# Analyses on Wassenius' Report for Total Solar Eclipse in 1733: Quantifications of the Solar Radius and the Earliest Reported Prominences

Hisashi Hayakawa (1-4), Mitsuru Sôma (5), Noortje Peek (6),

Jean-Pierre Rozelot (7), Stanislav Gunár (8), Alexei Pevtsov (9)

(1) Institute for Space-Earth Environmental Research, Nagoya University, Nagoya, 4648601, Japan

(2) Institute for Advanced Research, Nagoya University, Nagoya, 4648601, Japan

(3) Space Physics and Operations Division, RAL Space, Rutherford Appleton Laboratory, Science and Technology Facilities Council, Harwell Oxford, Didcot, Oxfordshire, OX11 0QX, UK

(4) Nishina Centre, Riken, Wako, 3510198, Japan

(5) National Astronomical Observatory of Japan, Mitaka, 1818588, Japan

(6) University of Amsterdam, Amsterdam, 1012 VB, the Netherlands

(7) Université de la Côte d'Azur, Grasse, 06130, France

(8) Astronomical Institute of the Czech Academy of Sciences, 25165 Ondřejov, Czech Republic

(9) National Solar Observatory, Boulder, CO 80303, USA

**Abstract**

Total solar eclipses (TSEs) offer a unique opportunity to observe the solar atmosphere, detect limb phenomena, and accurately measure the solar radius. Following the TSE in 1733, Wassenius first reported the existence of prominences to the scientific community. Wassenius' original manuscript is held in the Royal Academy Archives of Sweden; this study translates his report and documents the associated source materials and local eclipse visibility. The solar radius ($R_\odot$) during the TSE in 1733 are 696250 ± 170 km and 959.99 ± 0.24″ in the absolute and apparent scales, respectively. This result contrasts with the modern standard (helioseismic) $R_\odot$ of 695780 ± 160 km and 959.34 ± 0.22″; however, it is consistent with the solar radius recorded in 1715. The observed prominences are located at +23.5 ± 22.5°, +66.5 ± 22.5°, and −68.5 ± 22.5° in the heliographic latitude. The appearance of prominences at such high latitudes contrasts with the sunspot butterfly diagram for 1725–1750, confirming 1733 as a solar minimum. These high-latitude prominences can potentially be attributed to the so-called 'polar rush' prominences that appear a few years after a solar minimum. If they are categorised as 'polar rush' prominences, the solar minimum must be re-dated to before 1733 May. Furthermore, the latitudes of at least two of the prominences reported by Wassenius enable their classification as quiescent prominences, suggesting the presence of a polarity inversion line in the polar regions in early 1733.

1. Introduction

When viewed from the Earth's surface, the apparent (angular) lunar radius appears to be extremely close to that of the Sun. This coincidence enables the Moon to completely block the solar photosphere, while showing the solar atmosphere and limb phenomena during total solar eclipses (TSEs) (Pasachoff, 2017).

Past TSE records occasionally offer unique insights into the past variability of the Sun and Earth. TSEs offer the most reliable ground-based methods to measure the solar radius and quantify its long-term variability as they minimise atmospheric effects, unlike other ground-based observations (Rozelot and Damiani, 2012). TSE records have been actively used to study long-term variations in the solar radius in the 20th century; a few cases were revisited in the late 19th century (Dunham *et al*., 1980; Fiala *et al*., 1994; Rozelot and Damiani, 2012; Rozelot *et al*., 2018). The scientific community has analysed and discussed the isolated TSE in 1715, although some researchers debate if the solar radius at that time was larger than it is now (Dunham *et al*., 1980; Parkinson *et al*., 1980; Rozelot and Damiani, 2012). Measurements of the solar radius based on eclipses are benefiting renewed interest thanks to better determinations of the lunar topogrphy data (Araki *et al*., 2009) and more accurate ephemeris data (Park *et al*., 2021).

For millennia, TSEs have left an imprint on human history (Stephenson, 1997; Hayakawa *et al*., 2025) and contributed to several observational discoveries such as prominences, Helium, and Einstein Effect (Orchiston *et al*., 2015; Pasachoff, 2017). TSEs have also contributed to the development of modern scientific analyses of solar coronal dynamics (Loucif and Koutchmy, 1989; Yeates *et al*., 2018; Habbal *et al*., 2021; Qu *et al*., 2022), solar radius measurements (Kubo, 1993; Fiala *et al*., 1994; Kilcik *et al*., 2009; Lamy *et al*., 2015; Quaglia *et al*., 2021), and Earth rotation speed measurements (Stephenson, 1997; Stephenson *et al*., 2016). Solar prominences were first documented and published in scientific literature after the TSE in 1733 by Wassenius (1733). His report is considered the earliest robust record of solar prominence (Pettit, 1943; Vaquero and Vázquez, 2009; Charbonneau, 2014). Besides, there are some possible cases of prominence witnessed before this eclipse have been reported (Stephenson, 1997; Tandberg-Hanssen, 1998). For example, there is a description of possible prominences during the 1185 total solar eclipse in the Novgorod Chronicles, while some have associated this description with a chromosphere (Sviatsky, 1923; Stephenson, 1997; Pevtsov *et al*., 2016). Only in 1851, scientists began understanding this phenomenon (Tandberg-Hanssen, 1998), and from the early 20th century, these data were digitally available (Chatterjee *et al*., 2020).

Wassenius (1733) did not publish his prominence drawings, retaining a record of historical interest. His unpublished manuscript, which is held in the Royal Academy Archives of Sweden, provides further observational details beyond that reported in the literature (Wassenius, 1733) and his prominence drawings (Nordenmark, 1938). Wassenius' reports for the TSE in 1733 offer unique datasets composed of past solar activity. The scientific community has no data on the latitudinal distribution of prominence in the 18th century. The 1730s are known as fault lines for sunspot number calibrations and solar-cycle reconstructions because sunspot data availability was extremely scarce during this period (Hayakawa *et al*., 2022; Clette *et al*., 2023). The TSE that Wassenius reported can fill the chronological data gap between the 1715 and modern TSE reports in terms of solar radius measurements.

Therefore, this study aims to document his reports (Section 2) and local eclipse visibility (Section 3), measure the solar radius (Section 4) and prominence distributions (Section 5), and contextualise them into a larger picture (Section 6).

**2. Wassenius and his Records**

Birger Wassenius (1687–1771) joined Olof Hjorter's observational campaign on the TSE in 1733 May in Göteborg (Sweden)[1]. He was well known for the earliest unambiguous prominence report on his eclipse observation in 1733 (Charbonneau, 2014). At the time of this solar eclipse, Wassenius

---

[1] Wassenius was born in Mankärr near Vänersborg (Sweden) in 1687, studied mathematics, physics, and astronomy from 1712 to 1722 in Uppsala (Sweden) and later became a lecturer at a gymnasium of Göteborg until his retirement to Mankärr.   He was well known for the earliest unambiguous prominence report on his eclipse observation in 1733 (Charbonneau, 2014). At the time of this solar eclipse, Wassenius was a mathematics lecturer in Göteborg (Wassenius, 1733, p. 134).

was a mathematics lecturer in Göteborg (Wassenius, 1733, p. 134).

Wassenius' correspondence on this TSE held at the Royal Society Archives is written in Latin and catalogued as MS EL/V/67; it has been subsequently printed in *Philosophical Transactions* (Wassenius, 1733). His report held at the Royal Academy Archives of Sweden is mostly written in Swedish with drawings of probable prominence and catalogued as MS Wassenius. This report was subsequently transcribed into modern Swedish in Nordenmark (1938). The appendix shows our English translation of MS Wassenius, providing more details compared to MS EL/V/67 (Wassenius, 1733), including two sketches on probable prominences (Figure 1) yet to be analysed in the scientific community (Nordenmark, 1938).

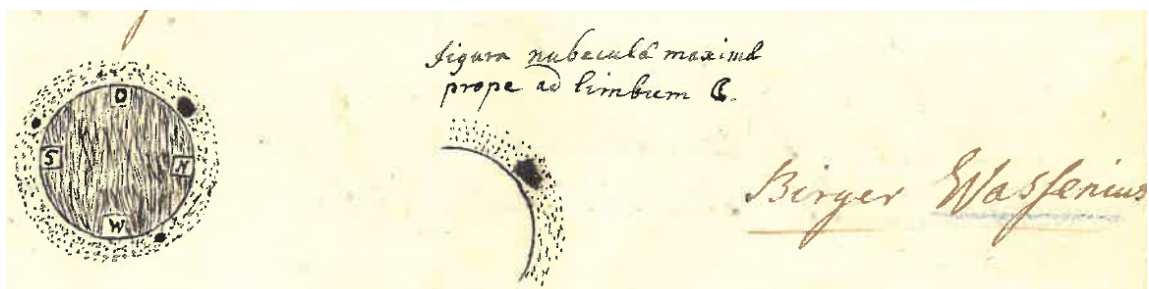

Figure 1: Wassenius' sketches of the eclipsed Sun and prominences on 1733 May 13, as reproduced from MS Wassenius (f. 2b), with a courtesy to the Center for History of Science of the Royal Swedish Academy of Sciences.

3. Local Eclipse Visibility

Wassenius dated this TSE to 1733 May 2. This date followed the Julian Calendar, which Sweden continued to use until 1748. According to the Gregorian Calendar, this eclipse was dated 1733 May 13. Wassenius was located in Cronan in Göteborg during the eclipse observations in 1733 (MS Wassenius, f. 1a); this has been identified as Skansen Kronan (N57°41′46″, E011°57′19″) located on the southwestern side of the old town area of Göteborg (Petersen, 2009, p. 35)[2].

---

[2] Wassenius used a 21-feet telescope to see the TSE (Wassenius, 1733, p. 135). He saw 'all the large brighter or darker spots of the Moon' and 'the light ring or the Corona around' from 'the tube in question [NB his telescope], very smooth/homogeneous and without streaks' (MS Wassenius, f. 1b). This implies that he did not project the solar image to the projection sheet but directly viewed the

Wassenius reported the local beginning and end of this TSE as 19:14:46 and 19:16:54 in the local apparent time (LAT), a total duration of 128 s. At 19:15:50 LAT, he reported the visibility of 'all the stars in Ursa Major, Cor Leonis, Procyon, Sirius, and a few more' (MS Wassenius, f. 1a). Wassenius measured the time with one-second precision using his wall clock and four other small clocks set and corrected according to the solar altitude (MS Wassenius, f. 1a). Thus, Wassenius used an 'eye-and-ear method' using pendulum clocks, which provide beats as short as 0.5 s with different beats, a method implemented by astronomers in the 18th century (Bell, 2025).

This is broadly consistent with the local eclipse visibility calculated developing the Sôma method (Sôma and Tanikawa, 2015). Here, we utilises the ephemeris data of NASA JPL DE 441 (Park *et al.*, 2021) and $\Delta T$ (offset between universal and terrestrial times) of 14 s (Stephenson *et al.*, 2016) without considering the lunar-limb profile. Based on our calculations, the eclipse reached the maximum at 19:15:42 LAT with a magnitude of $1.011^3$, lasting for 134 s from 19:14:35 LAT to 19:16:49 LAT; this was ≈ 6 s longer than the reported duration.

**4. Solar Radius of 1733**

Wassenius' report indicates a local totality duration of 128 s at the observational site at Skansen Kronan (N57°41′46″, E011°57′19″), which is 6 s shorter than the calculated value. These details enable calculating the solar radius ($R_\odot$) using the ephemeris data, Earth's rotation speed parameter, and lunar-limb profile. We compute the local eclipse visibility through the Sôma method (Sôma *et al.*, 2012) using the lunar topography data of the KAGUYA Mission (Araki *et al.*, 2009).

---

solar disk through the eyepiece. Owing to his observational method, 'no whole Moon could fall into my eyes [NB Wassenius' eyes] at once through such a long tube' (MS Wassenius, f. 2a). Wassenius concentrated on observing the lunar limb and prominences without solar filters because he mentioned a shot of the solar glare to his eye upon the end of the eclipse totality, probably upon the diamond ring or Baily's beads (MS Wassenius, f. 2a). Seemingly, Wassenius' field of view was limited to the neighbourhood of the lunar limb and too narrow to cover the wider coronal structure. This is probably why Wassenius missed outer solar coronal features.

[3] This study has defined the magnitude of eclipses (M) as M = (Sun's apparent angular semidiameter + Moon's apparent angular semidiameter – apparent angular distance of the centers of the Sun and Moon) / (Sun's apparent angular diameter), following Hayakawa et al. (2022).

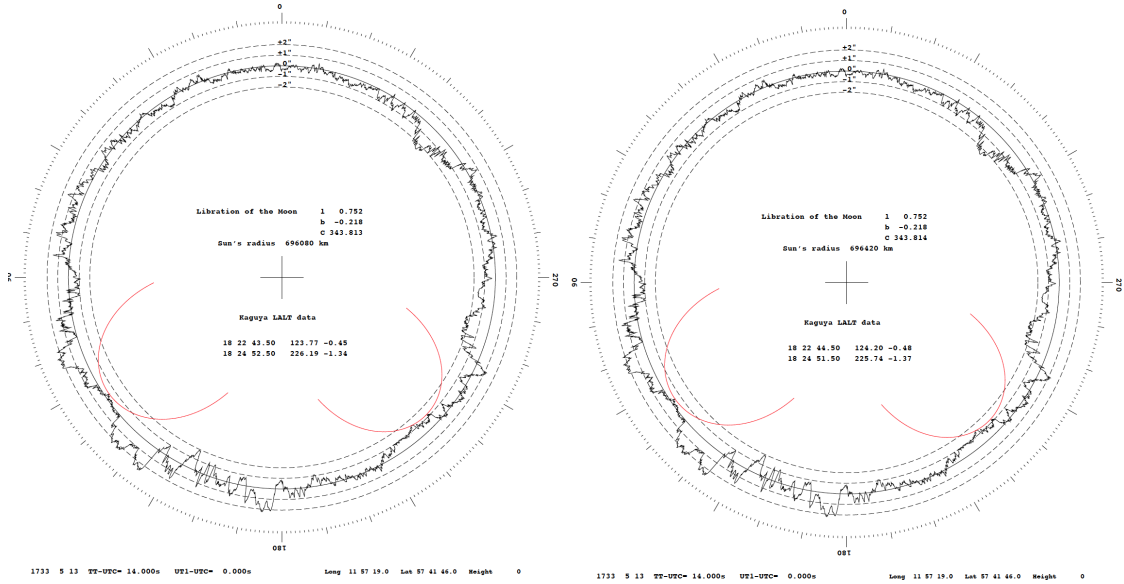

Figure 2: Lunar-limb profiles in comparison with different solar-limb profiles, setting $R_\odot$ = 696080 km (left) and $R_\odot$ = 696420 km (right), where the solar photosphere touches the bottom of the deepest lunar valleys. The position angles are measured from the direction of celestial north. The lunar topography data are derived from KAGUYA Mission (Araki *et al*., 2009), following the method of Sôma *et al*. (2012). The lunar-limb profile is emphasised by a factor of ≈ 50 for visual presentation; the solar limbs at the beginning and end of the totality are drawn by red curves with respect to the emphasised lunar limb.

Wassenius reported a TSE duration of 128 s in MS Wassenius and Wassenius (1733). Wassenius used five clocks for timekeeping and corrected them based on the solar altitude. We expected Wassenius' duration measurement accuracy to be ±1 s. Figure 2 shows totality duration with different $R_\odot$ values in comparison with the KAGUYA lunar topogtaphy data (Araki *et al*., 2009) following the method proposed by Sôma *et al*. (2012). Any excesses of the solar photosphere above the lunar limb are observed as Bailey beads. We need to set the $R_\odot$ margin of 696250 ± 170 km to satisfy the totality duration of 128 ± 1 s. This is slightly larger than the old canonical $R_\odot$ of 696000 km (Auwers, 1891; Archinal *et al*., 2011), modern standard (helioseismic) value of 695780 km (Takata and Gough, 2024) and nominal value of 695700 km (Prša *et al*., 2016).

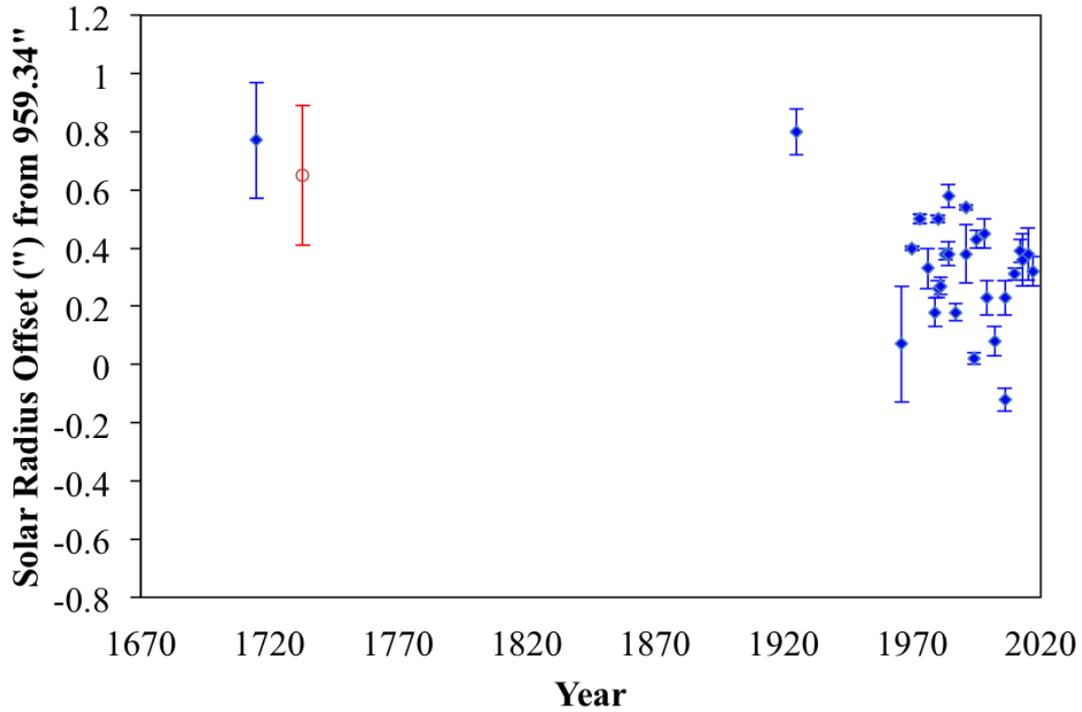

Figure 3: Eclipse-based $R_\odot$ measurements in comparison with the modern standard $R_\odot$ (959.34″; Takata and Gough, 2024), where our $R_\odot$ constrain is indicated by the red bar and the $R_\odot$ constrains of the previous studies by blue bars (Kubo, 1993; Kilcik *et al*., 2009; Lamy *et al*., 2015; Quaglia *et al*., 2021).

During this TSE, the Moon was ≈ 357778 km away from the observational site according to calculations performed using the Sôma method (Sôma *et al*., 2012) and NASA JPL DE 441 (Park *et al*., 2021). For a long-term comparison, our $R_\odot$ value was translated to 959.99 ± 0.24″ in the apparent $R_\odot$, setting the Sun at 1 au from the Earth. Subsequently, we calculated the departure from the modern standard $R_\odot$ = 959.34″ (Takata and Gough, 2024), which appears consistent with two other ground-based measurements (Rozelot *et al*., 2003; Morand *et al*., 2011).

The $R_\odot$ departure of 1733 is calculated as 0.65 ± 0.24″. Figure 3 compares our result with modern $R_\odot$ measurements reported in previous studies (Kubo, 1993; Kilcik *et al*., 2009; Lamy *et al*., 2015; Quaglia *et al*., 2021). This value is slightly larger than the modern standard $R_\odot$ values; however, it is consistent with the 1715 $R_\odot$ estimate of 0.77 ± 0.20″ (Dunham *et al*., 1980; Fiala *et al*., 1994; Kilcik *et al*., 2009) and remains in the variability range of modern measurements (Figure 3). This independently supports a long-term $R_\odot$ variability in the centennial timescale that contrasts the $R_\odot$ values of the early 18th century to the modern standard $R_\odot$ values.

## 5. Prominences

Wassenius gained recognition for his prominence observation in 1733. As shown in Figure 1, MS Wassenius included eclipse sketches with prominence locations he did not publish in Wassenius (1733). His eclipse sketches show unique positions of the three prominences on the eclipsed Sun. Wassenius showed solar orientations at four cardinal points: N, O, S, and W for the north, east, south, and west, respectively.

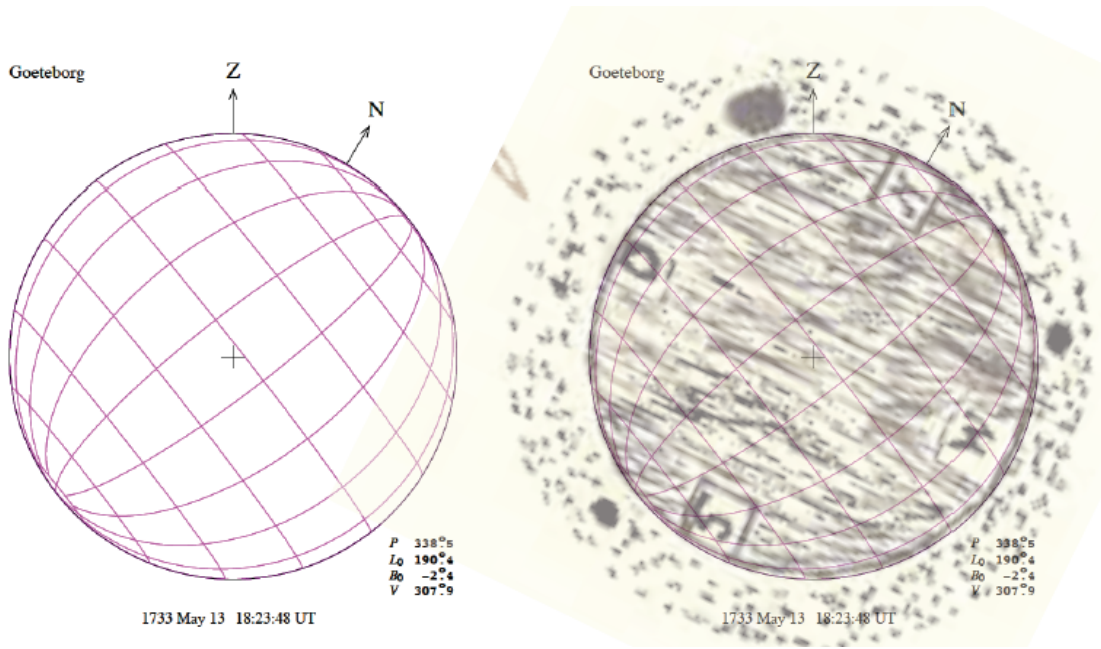

Figure 4: Heliographic coordinate of the eclipsed Sun at the maximum of the TSE in 1733, as seen from Skansen Kronan (left), in comparison with the overlay on Wassenius' eclipse sketch (right). We annotate the local zenith as Z and the local celestial north as N, respectively.

Using these cardinal points, we computed the apparent solar orientation as seen from Skansen Kronan (N57°41′46″, E011°57′19″). Figure 4 shows our calculation for the local orientation of the solar disc at the maximum of this TSE using the method by Hayakawa *et al*. (2021), ephemeris data of NASA JPL DE 441 (Park *et al*., 2021), and ΔT value of 14 s (Stephenson *et al*., 2016). Our calculations showed the solar rotation axis inclined by 52.1° clockwise from the local zenith during the entire phase. Our calculation yielded $B_0$[4] and $P$ angles[5] of 2.4° and 21.5° (clockwise), respectively. The solar rotation axis was inclined 21.5° clockwise from the celestial north upon totality.

We overlaid calculated heliographic coordinates on the Wassenius' eclipse drawing, following annotations made by Wassenius for disc orientations. However, slight caution is warranted for the reliability of the prominence positions in these sketches. Wassenius explicitly stated, 'no whole Moon could fall into my eyes at once through such a long tube' (MS Wassenius, f. 3). Wassenius 'had particular regard for the largest of these spots [NB: prominences], which in the tube was seen on the northeast side of the Moon' (MS Wassenius, f. 3). This is probably why Wassenius added the second sketch for the prominence on the northeast side besides his whole-disc sketch of the eclipsed Sun. We accommodate a positional uncertainty of ±22.5° because Wassenius described his positions in the eight cardinal directions. Wassenius' text enables locating this prominence at +23.5 ± 22.5° in the heliographic latitude (Figure 5), which is consistent with our measurement of Wassenius' eclipse sketch that located this prominence at ≈ +27°.

In contrast, Wassenius was less certain about the other two prominences. Wassenius himself recollected, 'In the hurry, I [NB: Wassenius] forgot to look for the other spots [NB: prominences]: though I remember that in the tube there were 2 quite small ones to be seen on the southern side of the Moon, and either one alone, or 2, on the lower or northwestern side' (MS Wassenius, f. 3). This description indicates that Wassenius did not record the positions of these two prominences in his eclipse sketch. Consequently, it can be assumed that Wassenius' sketch approximated the prominence position based on his recollection and was not more accurate than his textual description. This is also likely true for the altitude of prominences indicated in the drawings. The altitude seems to be above 0.2 $R_\odot$, which is high for typical quiescent prominences. However, it is extremely unlikely that Wassenius would have encountered eruptive prominences, and the lack of accuracy in drawing the prominences from recollection is a more likely explanation.

Wassenius located these prominences on the southern and northwestern sides. We assumed positional uncertainties of ±22.5° for each case and located the second and third prominences at +66.5 ± 22.5° and −68.5 ± 22.5°, respectively (Figure 5). The error margin may have been larger than our estimate because Wassenius only recollected their rough positions, missing the entire circumstances of the eclipsed Sun in his field of view. However, it is likely that at least one of the reported prominences was located at a high latitude, which is significant for understanding the phases of the solar cycle during the TSE in 1733. This is because high-latitude prominences are more likely to appear during the later parts of the rising phase of the solar cycle. This likelihood is documented in Chatterjee *et al.* (2020), who studied the concentrations of the appearances of prominences at high latitudes referred to as the 'polar rush' of prominences. Figures 5 and 7 in their paper show that these concentrations last for a few years, reaching their highest latitudes around the time when the Greenwich Sunspot Area values reach their (plateaued) maxima. After the 'polar rush' concentrations disappear, they reappear only during the later part of the rising phase of the next cycle. These figures also confirm high-latitude prominences outside of the 'polar rush' concentrations. The fraction of high-latitude prominences in the 'polar rush' concentrations can be estimated to be ≈ 80%.

---

[4] We use the $B_0$ angle to describe a heliographic latitude of the apparent central point of the Sun.

[5] We use the $P$ angle to describe a position angle of the rotation axis of the Sun with respect to the celestial north.

Wassenius was probably the first to describe the substructure of prominences. In his description, the largest of the reported prominences was 'composed of 3 reddish cloud cones placed diagonally to each other, with darker colours or stripes between, as the figure below somewhat shows', while his figure does not depict them. This description may indicate that the prominence was composed of several column-like or funnel-like parts, which are typical forms of quiescent prominence structures visible in lower-resolution optical observations. Such prominence substructures are occasionally referred to as 'prominence tornadoes'; however, they do not share any resemblance to tornadoes beyond the similarity of their silhouettes (Gunár *et al.*, 2023).

## 6. Summary and Discussions

This study examined Wassenius' unpublished manuscript in the Royal Academy Archives of Sweden for his TSE observation in 1733, in comparison with his report to the Royal Society of London (Wassenius, 1733). Wassenius' observation is a historical benchmark. He is associated with the earliest robust prominence observation in the scientific community. This manuscript includes the unique sketches of prominences around the eclipsed Sun (Figure 1). The features represent prominences, as Wassenius (1733) mentions their reddish colour and given that they were located towards the edge of the Moon but did not extend to Moon's surface. The features were observed for some extended time ('I looked again after, and could see as before, for 40 or 50 seconds'), and were not as bright as a 'flash', which he mentions in his description ('the Sun's beam shot into my eye like a rocket').

His records enable us to locate the observational site at Skansen Kronan (N57°41′46″, E011°57′19″) on the southwestern side of the old town area of Göteborg. Wassenius used a 21-foot telescope to observe the TSE. This instrument afforded him a close-up view of the detailed structure on the surface of the Moon and the eclipsed Sun with a loss of the broad overview of the entire eclipsed Sun. This is why he probably missed the coronal structure of the eclipsed Sun in 1733. Wassenius' drawing shows a ring of 'corona' up to ≈ 1.3 – 1.4 $R_\odot$, even if we take his drawings at face value. At these low heights, resolving the coronal structure is difficult. His drawings do not show decreases in coronal appearance in the direction away from solar limbs or in different areas around the disc (poles vs. equatorial regions). However, he noted a decrease in brightness further away from Moon edge ('fading more and more, the further one got out from the periphery of the Moon, but being on the more lit side as the Sun was'), which suggests that what he saw as the corona may have largely been scattered light. According to our calculations, the TSE occurred late in the evening, when the Sun was ≈ 6° above the horizon, supporting the hypothesis about the scattered light.

Wassenius reported the duration of the total solar eclipse as 128 s at Skansen Kronan (N57°41′46″, E011°57′19″). We set the $R_\odot$ margin of $R_\odot$ = 696250 ± 170 km and 959.99 ± 0.24″ in the absolute and apparent scales, respectively, to satisfy the totality duration of 128 ± 1 s. Our result contrasts with the modern standard (helioseismic) $R_\odot$ of 695780 km and 959.34 ± 0.22″ in the absolute and apparent scales, respectively. This value is slightly larger than the modern standard $R_\odot$; however, it is broadly consistent with the 1715 $R_\odot$ estimate of 0.78 ± 0.20″ (Figure 3). However, this variation was only 0.04 ± 0.03% above the modern standard value and hardly substantial, missing almost two centuries between the 1733 and 1925 eclipses. Our result does not contradict the alleged long-term $R_\odot$ decrease that Dunham *et al*. (1980) derived as −0.0013 ± 0.0008 arcsec/yr from three solar eclipses in 1715, 1976, and 1979, which Hiremath *et al*. (2020) supported based on a century of solar-radius measurements from the Kodaikanal Observatory. More data are required for confirmation.

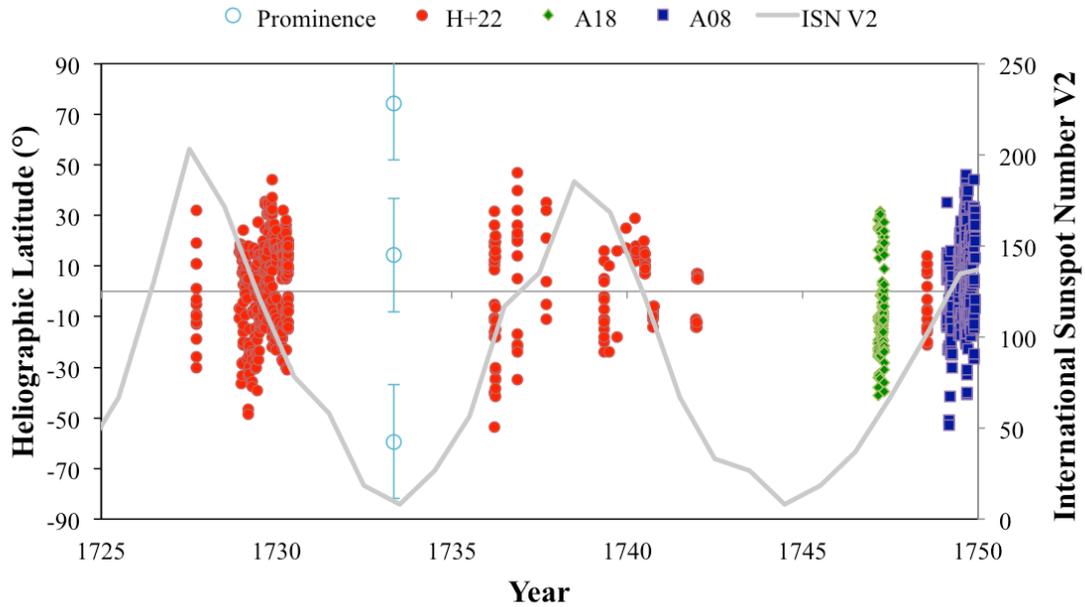

Figure 5: Our measurement for the heliographic latitudes of Wassenius' prominences (bars in light blue), in comparison with the contemporaneous sunspot position datasets (Arlt, 2008, 2018; Hayakawa *et al.*, 2022) and the International Sunspot Number V2 (ISN V2: Clette *et al.*, 2023). We need to be cautious on the reliability of the International Sunspot Number V2 before 1749, as the recent studies suggested substantial revisions in this period (Hayakawa *et al.*, 2022; Carrasco *et al.*, 2024).

We also examined Wassenius' report for the TSE in 1733 and the drawings locating prominences at +23.5 ± 22.5°, +66.5 ± 22.5°, and −68.5 ± 22.5° in the heliographic latitude. Despite the significant uncertainties, the appearance of prominences at such high latitudes contrasts with the sunspot butterfly diagrams for 1725−1750 (Hayakawa *et al.*, 2022) and the International Sunspot Number V2 (Clette *et al.*, 2023), which suggest 1733 was a solar minimum (Figure 5). If true, any sunspot groups in 1733 would have been close to the equator (Hathaway, 2015), the polar field would have developed completely, and there would be no, or at best, few, high-latitude prominences. If prominences that Wassenius reported were part of the prominence 'polar rush' suggested by the higher numerical probability of occurrence of high-altitude prominences (Chatterjee *et al.*, 2020), the solar cycle would have been in the late ascending phase. In this case, Solar Cycle −1 would have started considerably before 1733 May to develop the 'polar rush' of prominences, in contrast with the current understanding in the International Sunspot Number V2 (Clette *et al.*, 2023). However, quiescent prominences appear at high altitudes at random stages of the solar cycle with non-negligible likelihood (Chatterjee *et al.*, 2020). This probability can be estimated to be ≈ 20%. Thus, the appearance of the prominence at high altitudes indicates the presence of a polarity inversion line in the polar region in early 1733.

**Funding**

This research was conducted under the financial support of JSPS Grants-in-Aids JP25K17436 and JP25H00635, the ISEE director's leadership fund for FYs 2021 – 2025, the Young Leader Cultivation (YLC) programme of Nagoya University, Tokai Pathways to Global Excellence (Nagoya University) of the Strategic Professional Development Program for Young Researchers (MEXT), the young researcher units for the advancement of new and undeveloped fields in Nagoya University Program for Research Enhancement, Transdisciplinary Network linking Space-Earth Environmental Science, History, and Archaeology (JPMXP1324134720) of MEXT Promotion of Development of a Joint Usage/Research System Project: Coalition of Universities for Research Excellence Program (CURE), and the NIHU Multidisciplinary Collaborative Research Projects NINJAL unit 'Rediscovery of Citizen Science Culture in the Regions and Today'. The U.S. National Solar Observatory is operated by the Association of Universities for Research in Astronomy (AURA), Inc. under a cooperative agreement with the National Science Foundation. S.G. acknowledges the support from grant 25-18282S of the Czech Science Foundation (GAČR).

**Data Availability**

The MS Wassenius is located in the Center for History of Science of the Royal Swedish Academy of Sciences. The International Sunspot Number V2 is provided from the Sunspot Index and Long-term Solar Observations (SILSO) of the Royal Observatory of Belgium. The ephemeris data are provided from Park *et al*. (2021). The ΔT data are provided from Stephenson *et al*. (2016). The lunar topography data are provided from the KAGUYA Mission (Araki *et al*., 2009).

**Acknowledgments**

We wish to thank the Center for History of Science of the Royal Swedish Academy of Sciences for letting us access MS Wassenius. We wish to thank the Royal Society Archives of London for letting us access MS EL/V/67. HH thanks Yukiko Kawamoto for her help on the interpretation and translation of Wassenius (1733), Frédéric Clette for his insightful advice on interpreting Wassenius' coronal descriptions, and Marinus A. van der Sluijs for his help to have a contact with Nordom. Eclipse calculations were partly carried out on the Multi-wavelength Data Analysis System operated by the Astronomy Data Center, National Astronomical Observatory of Japan.

**Appendix: Translation of Wassenius' Report to the Royal Academy of Sweden (MS Wassenius)**

Solar Eclipse dated 2 May 1733 observed in/from Göteborg, at a castle, named Cronan[6], approximately 1,000 al:[7] from the city itself.

The time, according to my clock which shows and clicks minutes and seconds, as well as 4 other smaller ones, set and duly corrected after the height of the Sun.

The rise and fall of the eclipse, etcetera.

Hour. Minute. Second:

6. 26. 0. It was noticed that the eclipse had already begun, in the northwest, the edge of the Sun: after the very beginning, it was not possible to determine it precisely due to clouds.

6. 38. 41. The Sun seemed to be eclipsed for 3 digits approximately.

6. 49. 52. The Sun was eclipsed for 6 digits approximately.

7. 14. 6. ㄹ was clearly visible, equally to Sirius.

---

[6] This is most probably associated with Skansen Kronan.

[7] This could be an abbreviation of 'altitude' or a similar word.

7. 14. 46. The Sun was completely covered, and immediately the Moon's atmosphere was clearly seen with the tube.

7. 15. 50. Equally bright on the northwestern and southeastern edges of the Moon; but a little brighter on the southwestern side: all the stars in Ursa Major, Cor Leonis, Procyon, Sirius, and a few more were seen; but nothing; no or Mars.

7. 16. 54. The Sun begins to light up again with great speed and the Moon's atmosphere disappears.

7. 20. 12. ♃ was still visible.

7. 41. 38. The Sun was still eclipsed, just to 6 digits.

8. 5. 50. The Sun became completely full, and the eclipse ended with the southeastern ⊙ edge; this is as much as one could deduct 55 or 56 degrees from a vertical line drawn through the ⊙ disc.

The Sun completely covered in Göteborg: 2 minutes 8 seconds.

As soon as the Sun here in Göteborg was completely obscured, I saw the true circumference of the Moon through my tube, which is approximately 21 feet long, very clearly. Just like all the large brighter or darker spots of the Moon. The light ring or the Corona around seemed, seen from the tube in question, very smooth/homogeneous and without streaks; fading more and more, the further one got out from the periphery of the Moon, but being on the more lit side as the Sun was. The most remarkable thing that I can tell the Royal Society is this; that immediately after the total obscuration of the Sun, there were some small brighter spots to be seen in the bright ring or atmosphere, about 3 or 4, of different temperament and size: which leaned in towards the periphery of the Moon, but in no place completely into the same. As now no whole Moon could fall into my eyes at once through such a long tube, I had particular regard for the largest of these spots, which in the tube was seen on the northeast side of the Moon. It was composed of 3 reddish cloud cones placed diagonally to each other, with darker colors or stripes between, as the figure below somewhat shows. I wondered about it, and stroked myself around and in the left eye, with which I looked; but I saw it the same way. Then left another to look there after: but as he did not immediately find the Moon, I looked again after, and could see as before, for 40 or 50 seconds. In the hurry, I forgot to look for the other spots: though I remember that in the tube there were 2 quite small ones to be seen on the southern side of the Moon, and either one alone, or 2, on the lower or northwestern side. I thought I would get a glimpse before the Sun could rise: but at the same time I was looking for them, the Sun's beam shot into my eye like a rocket: so I shouted that one should observe the timestamp carefully, etc.

That the spots mentioned must have been some clouds in our air: as Pastor in Marstrand Mr. Mag. Brag said to me, a few days after the eclipse, that they thought that, as they were two miles from here, with a beautiful English perspective, they would also have seen something like that around the Moon's circumference, that I can never believe; for our clouds do not appear so light and solid (:coherent bodies :) through my tube, but like a fog; nor did the clouds stand so still that day; but went quite fast, and disappeared or diminished towards the evening very quickly. In the eye or glass there is no cause, for this is subject to the illustrious Royal Society's own examination.

On Svenäcker rectory[8], a mile from Vänersborg in the south-southwest, under the Polaris' altitude 58 degrees 15 or 16 minutes, priest Torsten Wassenius observed the beginning of the total Solar eclipse at 7 17' 34" o'clock, the end of the total eclipse 7 20' 5"; so that the total eclipse lasted 2 minutes and 31 seconds.

The shape of the largest cloud near the ☽ limb

Birger Wassenius

---

[8] Although Nordenmark's transcription convincingly corresponds with the handwriting of Wassenius' original manuscript in most cases, doubt arises regarding his reading of "Stieråker" here. Torsten Wassenius was mentioned with the site name of 'Svenæcker (Svenäcker)' in his contemporary sources. For example, Celsius listed one of the auroral report as "On 1729 April 21 [in Julian Cakendar] in Svenæker near the Gothic waterfalls called Trollhætta, Torstanus Wassenius. V. D. Comminister in Wassenda (A. 1729 d. 21 April. in Svenæker prope Albis Gothici cataractas Trollhætta dictas, Torstanus Wassenius. V. D. Comminister in Wassenda)" (Celsius, 1733, p. 32). Such testimonies allow us to confidently correct Nordenmark's reading of 'Stieråker' to Svenäcker. We corrected his reading accordingly.